\documentclass[10pt,letterpaper]{IEEEtran}
\usepackage{amsmath}
\usepackage{graphicx}
\usepackage{color}
\usepackage{amsfonts}
\usepackage{float}
\usepackage{mathtools,algorithm}
\usepackage[noend]{algpseudocode}
\floatstyle{ruled}
\newfloat{algorithm}{tbp}{loa}
\providecommand{\algorithmname}{Algorithm}
\floatname{algorithm}{\protect\algorithmname}
\ifCLASSINFOpdf
\else
\fi

\begin{document}

\title{Designing Anti-Jamming Receivers for NR-DCSK Systems Utilizing ICA, WPD, and VMD Methods}
\author{Binh Van Nguyen, Minh Tuan Nguyen, Hyoyoung Jung, and Kiseon Kim
\thanks{Binh Van Nguyen is with the Institute of Research and Development, Duy Tan University,
Da Nang 550000, Vietnam. He is also with the School of Electrical Engineering and Computer Science,
Gwangju Institute of Science and Technology, Republic of Korea. (E-mail : binhnguyen@gist.ac.kr).}
\thanks{Minh Tuan Nguyen, Hyoyoung Jung, and Kiseon Kim are with the School of
Electrical Engineering and Computer Science, Gwangju Institute of Science and Technology, Republic
of Korea. (E-mail : {nguyenminhtuan,rain, kskim}@gist.ac.kr).}
\thanks{The authors gratefully acknowledge the support from Electronic Warfare Research Center at
Gwangju Institute of Science and Technology, originally funded by Defense Acquisition Program
Administration (DAPA) and Agency for Defense Development (ADD).}
}
\maketitle
\begin{abstract}
In this work, we consider an advanced noise reduction differential chaotic shift keying
(NR-DCSK) system in which a single antenna source communicates with a single antenna
destination under the attack of a single antenna jammer. We devote our efforts to design
a novel anti-jamming (AJ) receiver for the considered system. Particularly, we propose a
variational mode decomposition-independent component analysis-wavelet packet
decomposition-based (VMD-ICA-WPD-based) structure, in which the VMD method is
firstly exploited to generate multiple signals from the single received one. Secondly, the ICA
method is applied to coarsely separate chaotic and jamming signals. After that, the WPD
method is used to finely estimate and mitigate jamming signals that exist on all outputs of
the ICA method. Finally, an inverse ICA procedure is carried out, followed by a summation,
and the outcome is passed through the conventional correlation-based receiver for recovering
the transmitted information. Simulation results show that the proposed receiver provides
significant system performance enhancement compared to that given by the conventional
correlation-based receiver with WPD, i.e. $8$ dB gain at BER $=0.03$ and Eb/N0 $= 20$ dB.
\end{abstract}

\begin{IEEEkeywords}
Chaotic communication, anti-jamming receiver, independent component analysis,
wavelet packet decomposition, variational mode decomposition, bit-error-rate.
\end{IEEEkeywords}

\IEEEpeerreviewmaketitle

\section{Introduction}
As an alternative to pseudo-random noise (PN) sequences used in conventional
spread spectrum systems, which have poor cross-correlation and low security
level, chaotic communication (CC) is proposed in \cite{Kaddoum-16}.
The basic idea of CC is to replace PN sequences by chaotic sequences. Chaotic
systems can be categorized as coherent and non-coherent ones, among which
the non-coherent chaotic systems have been widely considered and investigated
in the literature due to its simplicity \cite{Kaddoum-16}. In addition, among
various non-coherent chaotic systems, differential chaotic shift keying (DCSK)
is the fundamental and most studied one. The performance of the DCSK system
under the additive white Gaussian noise (AWGN), multipath fading, and single-tone
jamming environments are extensively investigated in \cite{Long-11},
\cite{Xia-04}, and \cite{Lau-02}.

Although the DCSK scheme is largely investigated, it has several drawbacks, i.e.
low data rate, high energy consumption, and high complexity. To alleviate these
drawbacks, several advanced alternatives have recently proposed. Particularly,
high efficiency DCSK (HE-DCSK) \cite{Yang-12}, improved DCSK (I-DCSK)
\cite{Kaddoum-15}, and short reference DCSK (SR-DCSK) \cite{Kaddoum-16-SRDCSK}
schemes are proposed to improve the data rate and enhance the spectral efficiency
of the DCSK scheme. In addition, in  \cite{Yang-13}, reference modulated DCSK
(RM-DCSK) scheme is presented to improve the transmission reliability and reduce
the complexity of the DCSK scheme. Moreover, in \cite{Kaddoum-16-NRDCSK},
noise reduction DCSK (NR-DCSK) scheme is introduced to reduce the noise variance
present in the received signal, and thus, enhance the system performance. Among the
aforementioned schemes, NR-DCSK provides the best performance. Recently, the
effects of several common jamming types, i.e. broad-band, partial-band, tone, and
sweep jamming, on the performance of the NR-DCSK system are extensively
investigated in \cite{Nguyen-18}.

The performances of the DCSK and the NR-DCSK systems under jamming
environments are studied in \cite{Lau-02} and \cite{Nguyen-18}, however,
how to mitigate the effects of jamming on the performances of chaotic systems
is still an open question. Since the inherent AJ capabilities of chaotic systems
may not work effectively when the jamming power is much larger than that
of chaotic signals, designing additional jamming mitigation techniques for
chaotic systems is of paramount importance. One of common practices to
mitigate jamming signals in wireless communication systems is to use fixed or
adaptive north filters (NFs) \cite{Ghatak-11}-\cite{Regalia-10}. While fixed
NFs do not perform well if jamming frequency is not known or time variant,
adaptive NFs incur high complexity because they use multiple sampling
frequencies and adaptively weight filters to identify jamming frequencies. If
multiple receiving antennas are available, jamming signals can be mitigated
through nulling \cite{Bhunia-16}. However, in this case, direction of arrivals
of jamming signals must be estimated first, which leads to high cost and
computational complexity. The other simple and promising alternatives
are to use independent component analysis (ICA) \cite{Raju-06}-\cite{Ranjith-16}
and wavelet packet decomposition (WPD) methods \cite{Mosavi-14}-\cite{Mosavi-16}.
The advantage of the ICA method lies in the fact that it does not require
prior information of jamming signals, i.e. jamming frequency and type. In
addition, the WPD method provides a detailed representation of a signal
in the time-frequency domain which is greatly useful for estimating and
mitigating jamming signals.

We observe that although the ICA and the WPD methods can be readily applied
to chaotic systems and may provide good AJ performance, there are still spaces
for improvement. Particularly, since the ICA method cannot tell which output is
the desired signals, additional resources are required to transmit pilots together
with the legitimate signal for the purpose of signal identification at the destination.
In addition, the ICA method is not applicable for a system having a single-antenna
destination because it requires that the number of input signals is at least equal
to the number of independent sources. Moreover, the ICA and the WPD methods
can also be combined together to enjoy advantages of both methods for jamming
estimation and mitigation.

Motivated by the aforementioned observations, in this
work, we consider the recent advanced NR-DCSK system, in which a single
antenna source communicates with a single antenna destination under the attack
of a single antenna jammer, and design a novel AJ receiver for it. Specifically,
we propose a novel correlation-based receiver with variational mode
decomposition-ICA-WPD (VMD-ICA-WPD) structure, in which we exploit the
VMD technique \cite{Nguyen-17-VMD} to generate multiple signals from the
received one, from which the ICA and the WPD methods can then be applied.
In addition, in the proposed receiver, we apply the WPD method on all outputs
of the ICA method, then an inverse ICA procedure and a summation are carried
out to obtain an estimation of the transmitted signal. This means that we do not
need to identify which output of the ICA method is which, and thus, leads to
system complexity reduction. We extensively simulate the AJ performance of
the proposed receiver and show that it significantly outperforms conventional
counterparts.

\section{System Model}
We consider the NR-DCSK system consisting of a source $\mathrm{S}$, a
destination $\mathrm{D}$, and a jammer $\mathrm{J}$. Each node is equipped
with a single antenna. While $\mathrm{S}$ is sending legitimate information to
$\mathrm{D}$, $\mathrm{J}$ tries to disrupt $\mathrm{D}$'s reception by
emitting jamming signals.

In the NR-DCSK system, the transmitted signal of the $l^{th}$ bit, $b_l$, can be
expressed as \cite{Kaddoum-16-NRDCSK}
\begin{align}
{s_k^l} = \left\{ \begin{array}{l}
{x_{\left\lceil {\frac{k}{P}} \right\rceil}^ l}, \;\;\;\;\;\;\;\; \text{if} \;\; 0 < k \le \beta, \\
{b_l}{x_{\left\lceil {\frac{k}{P}} \right\rceil  - \beta }^l}, \; \text{if} \;\; \beta  < k \le 2\beta,
\end{array} \right.
\end{align}
where ${\left\lceil {\cdot} \right\rceil }$ denotes the ceiling operator. In other words,
each bit duration is divided into two equal time slots, one for transmitting a reference
sequence of length $\beta$ and one for sending an information-bearing (IB) sequence.
To generate the reference sequence, a chaotic generator first generates $\beta/P$
chaotic samples. Then, each chaotic sample is replicated $P$ times. In addition, the
IB sequence is either the reference sequence (if the bit being sent is +1) or an inverted
version of the reference sequence (if the bit being sent is -1).

The received signal at the destination is given by
\begin{align}
r_k^l = h_{sd}^l s_k^l + h_{jd}^l j_k^l + n_k^l,
\end{align}
where $r_k^l$, $j_k^l$, and $n_k^l$ respectively denote the received signal,
jamming signal, and additive white Gaussian noise (AWGN) at the destination,
and $h_{sd}^l$ and $h_{jd}^l$ are the fading coefficients of the
source-destination and jammer-destination channels. The channels are assumed
to be independent block fading, which stay the same for a block of transmission
and independently change from one block to another.

We consider the sweep (chirp) jamming model which is shown to be the most
powerful jamming type against the NR-DCSK system \cite{Nguyen-18}. A
sweep jammer emits a narrow-band signal which has a time-varying frequency.
More specifically, the jamming signal sweeps from a starting frequency
$f_{start}$ to a stopping frequency $f_{stop}$ with a sweep rate $\Delta f$
during a sweep time $T_{sw}$. Among various sweep jamming variants, which
depend on narrow-band signals being used and sweep methods, the common
one based on sinusoidal signal and linear sweep method can be expressed as
follows \cite{Nguyen-18}
\begin{align}
j \left( t \right) = \sqrt {2{P_j}} \sin \left( {2\pi {f_{start}}t + \pi \Delta f {t^2} + \theta_{sw} } \right),
\end{align}
where $\Delta f = \frac{f_{stop} - f_{start}}{T_{sw}}$ and $\theta_{sw}$
denotes the initial phase of the sweep jamming signal. An equivalent discrete
baseband model of the above sweep jamming signal is expressed as
\begin{align}
j_k^l = \sqrt {2{P_j}} \sin \left( {\pi \frac{{k{F_{start}}}}{\beta } + \pi \frac{{{k^2}\Delta F}}{{4{\beta ^2}}}
 + {\theta _{sw}}} \right),
\end{align}
where $F_{start} = f_{start} T_b$, $\Delta F = \Delta f T_b^2$, and $T_b$ is the
bit duration.

\section{Conventional Receivers}
\subsection{Correlation-based Receiver \cite{Nguyen-18}}
\begin{figure}[!t]
    \centering
    \includegraphics[width = 8cm]{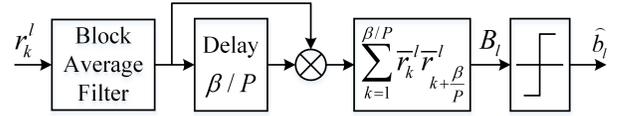}
    \caption{The correlation-based receiver for the NR-DCSK system \cite{Kaddoum-16-NRDCSK}.}
\end{figure}
The correlation-based receiver for the NR-DCSK system is illustrated in Fig 1. The received
signal $r_k^l$ is first fed into a moving average filter (MAF) which has a block size
of $P$. Secondly, the output of the MAF is correlated with its replica which is delayed
by $\beta/P$ samples, and then the outcome is summed over $\beta/P$ samples.
Finally, the sum is passed through a threshold detector to recover the transmitted
information. Mathematically, the output of the MAF can be written as
\begin{align}
\bar r_k^l = h_{sd}^ls_k^l + \frac{{h_{jd}^l}}{P}\sum\limits_{p = \left( {k - 1} \right)P + 1}^{kP} {j_p^l}
+ \frac{1}{P}\sum\limits_{p = \left( {k - 1} \right)P + 1}^{kP} {n_p^l},
\end{align}
where $\frac{{h_{jd}^l}}{P}\sum\limits_{p = \left( {k - 1} \right)P + 1}^{kP} {j_p^l}$ and
$\frac{1}{P}\sum\limits_{p = \left( {k - 1} \right)P + 1}^{kP} {n_p^l}$ are jamming signal
and noise after passing through the MAF. In addition, the decision variable $B_l$ is given by
{\small
\begin{align}
& {B_l} = \sum\limits_{k = 1}^{\beta /P} {\left( {h_{sd}^lx_k^l + \frac{{h_{jd}^l}}{P}\sum\limits_{p = \left( {k - 1} \right)P + 1}^{kP} {j_p^l}
+ \frac{1}{P}\sum\limits_{p = \left( {k - 1} \right)P + 1}^{kP} {n_p^l} } \right)} \nonumber \\
& \times \left( {h_{sd}^l{b_l}x_k^l + \frac{{h_{jd}^l}}{P}\sum\limits_{p = \left( {k - 1} \right)P + 1}^{kP} {j_{p + \beta }^l}
 + \frac{1}{P}\sum\limits_{p = \left( {k - 1} \right)P + 1}^{kP} {n_{p + \beta }^l} } \right).
\end{align} }
\subsection{Correlation-based Receiver with WPD}
Combining the WPD method \cite{Mosavi-16} and the correlation-based receiver readily gives a
correlation-based receiver with WPD, as shown in Fig. 2 where
$\mathord{\buildrel{\lower3pt\hbox{$\scriptscriptstyle\frown$}}\over j}$ denotes an estimation
of the jamming signal. $\mathord{\buildrel{\lower3pt\hbox{$\scriptscriptstyle\frown$}}\over j}$
is obtained by using the denoise capability of the WPD method, in which the jamming signal is
treated as the major one and the chaotic signal is considered as noise.

The basic idea of the WPD method is that it decomposes the input signal into
approximation and detailed parts, which are further split into other approximation and
detailed parts, and the process is repeated. The WPD method can also be interpreted as
a multi-resolution analysis, i.e. a $L$ level decomposition of the input signal into the
scaling and the wavelet functions, which are respectively expressed as
$\phi \left( t \right) = \sqrt 2 \sum\limits_k {h\left[ k \right]\phi \left( {t - k} \right)}$
and $\psi \left( t \right) = \sqrt 2 \sum\limits_k {g\left[ k \right]\psi \left( {t - k} \right)}$.
Consequently, the input signal can be reconstructed as
\begin{align}
r_k^l = \sum\limits_{m = 1}^L {\sum\limits_{n \in \mathbb{Z}} {{c_{m,n}}{\phi _{m,n}}} }  +
\sum\limits_{m = 1}^L {\sum\limits_{n \in \mathbb{Z}} {{d_{m,n}}{\psi _{m,n}}} },
\end{align}
where ${c_{m,n}} = \left\langle {r,{\phi _{m,n}}} \right\rangle$,
${d_{m,n}} = \left\langle {r,{\psi _{m,n}}} \right\rangle$,
${\phi _{m,n}} = {2^{ - m/2}}\phi \left( {{2^{ - m}}t - n} \right)$,
${\psi _{m,n}} = {2^{ - m/2}}\psi \left( {{2^{ - m}}t - n} \right)$, and $\left\langle {x,y} \right\rangle$
denotes the inner product between $x$ and $y$. $c_{m,n}$ and $d_{m,n}$ here are
called the wavelet coefficients.

For a denoising purpose, an appropriate threshold should be selected. Among various
options, the common threshold is as follows \cite{Mosavi-14}-\cite{Mosavi-16}
\begin{align}
\bar \gamma  = \delta \sqrt {2\ln \left( {N\ln \left( N \right)} \right)/\ln \left( 2 \right)},
\end{align}
where $\delta  = \mathrm{MAD} / 0.675$, $\mathrm{MAD}$ is the median absolute deviation
of the wavelet coefficients, and $N$ is the length of the input signal. Then, if a wavelet coefficient
exceeds $\bar{\gamma}$, we threshold it to zero. Thereafter, an estimation of the jamming
signal is obtained by applying the inverse WPD procedure on the modified wavelet coefficients.
Finally, the estimated jamming signal is removed from the received signal and the outcome is fed
into the conventional correlation-based receiver for recovering the transmitted information.
\begin{figure}[!t]
    \centering
    \includegraphics[width = 6cm]{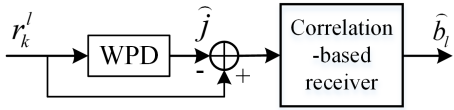}
    \caption{A correlation-based receiver with WPD for the NR-DCSK system.}
\end{figure}
\section{Correlation-based Receiver with VMD-ICA-WPD}
\begin{figure*}[!t]
  \centering
  \includegraphics[width = 11cm]{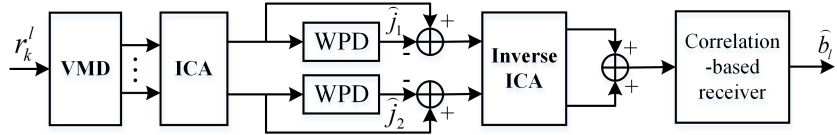}
  \caption{The proposed correlation-based receiver with VMD-ICA-WPD for the NR-DCSK system.}
  \vspace*{1pt}
  \hrulefill
\end{figure*}
In this section, we will propose a novel correlation-based receiver with VMD-ICA-WPD
which is suitable for a single-antenna destination and can provide advantages of both
the ICA and the WPD methods for the jamming estimation and mitigation purpose.
Particularly, we exploit the VMD method to generate multiple signals from the single
received one, after which the ICA and the WPD methods can be applied. The
proposed receiver is shown in Fig. 3, shown at the top of the next page.

The VMD method decomposes the given input signal into several band-limited
modes (narrow-band sub-signals) non-recursively. Each mode has an unique
center frequency (CF). The modes and their corresponding CFs are derived
by solving the following constrained variational problem
\begin{align}
\mathop {\min }\limits_{{v_n},{w_n}} &\;\;\;\;\;\;\;\;\;\; \sum\limits_n
{\left\| {{\partial _k}\left[ {\left( {\delta \left( k \right) + j/\pi k} \right) *
{v_n}\left( k \right)} \right]{e^{ - j{w_n}k}}} \right\|_2^2} \nonumber \\
\rm{s.t.} &\;\;\;\;\;\;\;\;\;\; \sum\limits_n {{v_n}}  =  r,
\end{align}
where $\left\| x \right\|_2^2$ is the $L^2$-norm of $x$, $\ast$ denotes the
convolution, $v_n$ and $w_n$ respectively are the $n^{th}$ mode and its
CF, $\partial_k$ denotes the gradient, $\delta \left( k \right)$ denotes the
impulse function, and $r = \left[ { \cdots r_k^l \cdots } \right]$ is a received
signal block. The problem is solvable by applying the augmented Lagrangian
multiplier method. The modes are directly updated in the Fourier domain. In
addition, each CF is re-estimated based on residuals of all other modes and
considered as the center of gravity of the power spectrum of the mode
\cite{Nguyen-17-VMD}. It is commonly suggested that the number of modes
should be large so that their summation can sufficiently resemble the input signal.
Following \cite{Nguyen-17-VMD}, we decompose the received signal into 10
modes, which are then divided into two mutually exclusive sets. After that,
members of each set are summed up to generate two signals, i.e. $V_1$
and $V_2$, which will be used as inputs of the ICA method. We use
only two signals as inputs of the ICA method to reduce its complexity.

The ICA method's inputs can be equivalently expressed as
\begin{align}
\mathbf{R} = \mathbf{H} \mathbf{X} + \mathbf{N},
\end{align}
where $\mathbf{R}= \left[ V_1 \;\; V_2 \right]^T$, $\mathbf{H}$ denotes
a possible mixing matrix,
$\mathbf{X} = {\left[ {\left( { \cdots s_k^l \cdots } \right) \;\; \left( { \cdots j_k^l \cdots } \right)} \right]^T}$,
and $\mathbf{N} = {\left[ {\left( { \cdots n_k^l \cdots } \right) \;\; \left( { \cdots n_t^l \cdots } \right)} \right]^T}$.
The ICA method computes an inverse of $\mathbf{H}$, i.e. $\mathbf{W}$, from
which coarse estimations of chaotic and jamming signals can be revealed by using
the equation $\tilde{\mathbf{X}} = \mathbf{WR}$. According to the Central Limit
Theorem, i.e. a sum of random variables tends to follow the Gaussian distribution,
the ICA method estimates the component of $\mathbf{W}$ in a way such that the
"non-Gaussianity" of each row of $\tilde{\mathbf{X}}$ is maximized \cite{Raju-06}.

In the next step, we apply the WPD method on all outputs of the ICA method to
estimate and mitigate jamming signals. Then an inverse ICA procedure is applied on
the de-jammed signals, i.e. multiplying the de-jammed signals with $\mathbf{W}^{-1}$,
and the outcomes are summed up to obtain an estimation, possibly jamming-free,
of the transmitted signal. It is noteworthy that the summation operation here actually
is equivalence to an inverse VMD procedure. Finally, the summed signal is fed into
the correlation-based receiver for recover the transmitted information bits. A
pseudo-code of the proposed receiver can be expressed as in Algorithm 1, shown
in the next page.
\begin{algorithm}
  \caption{Correlation-based receiver with VMD-ICA-WPD}
  \begin{algorithmic}[1]
    \Procedure{VMD}{} %\Comment{We have the answer if~$r$ is $0$}
      \State Input: a block of the received signal $r$
      \State Initialize $v_n^1$, $w_n^1$, $m \leftarrow 0$, $N = 10$ modes
      \While{not converged}
        \State Update $v_n$, $w_n$, and dual ascent $\lambda^m$
        \State $v_n^{m + 1} \leftarrow \mathop {\arg \min }\limits_{{v_n}} \mathfrak{L} \left( {v_{i < n}^{m + 1},v_{i \ge n}^{m + 1},w_i^m,{\lambda ^m}} \right)$
        \State $w_n^{m + 1} \leftarrow \mathop {\arg \min }\limits_{{w_n}} \mathfrak{L} \left( {u_i^{m + 1},w_{i < n}^{m + 1},w_{i \ge n}^m,{\lambda ^m}} \right)$
        \State ${\lambda ^{m + 1}} \leftarrow {\lambda ^m} + \tau \left( {r - \sum\nolimits_n {v_n^{m + 1}} } \right)$
      \EndWhile
      \State Output: ${V_1} = \sum\nolimits_{n = 1}^5 {{v_n}} $, ${V_2} = \sum\nolimits_{n = 6}^{10} {{v_n}}$.
    \EndProcedure
    \Procedure{ICA}{}
      \State Input: $\mathbf{R} = \left[V_1 \;\; V_2\right]^T$
      \State Centering: $\mathbf{R}_1 = \mathbf{R} - mean[\mathbf{R}]$
      \State Whitening: $\mathbf{R}_2 = \mathbf{C}^{ - 1/2} \mathbf{R}_1$, where $\mathbf{C} = E[\mathbf{R}_1 \mathbf{R}_1^H]$
     \For{p = 1:2}
        \While{not converged}
            \State Initialize $w_p$
            \State {\small $w_p^+ \leftarrow E\left[ \mathbf{R}_2 g\left( w_p^T \mathbf{R}_2 \right) \right] - E\left[ g'\left( w_p^T \mathbf{R}_2 \right) \right] w_p$}
            \State where $g(x) = x^3$ \cite{Shen-17}, $g^{'} (x)$ is its derivative.
            \State Decorrelation: $w_p^{++} \leftarrow w_p^{+} - \sum \nolimits_{i<p} w_i^T w_i w_p^{+}$
            \State Normalizing: $w_p = w_p^{++} /\left\| w_p^{++} \right\|$
        \EndWhile
      \EndFor
      \State Output: $\left[I_1 \;\; I_2\right]^T = \mathbf{W}^H \mathbf{R}$, where $\mathbf{W} = [w_1 \;\; w_2]^T$
    \EndProcedure
    \Procedure{WPD}{}
        \State Decompose $I_1$ and $I_2$ to $L$ detailed level.
        \State If a wavelet coefficient exceeds $\bar{\gamma}$, thresholds it to zero
        \State $\left[ {{{\mathord{\buildrel{\lower3pt\hbox{$\scriptscriptstyle\frown$}}
                \over j} }_1},{{\mathord{\buildrel{\lower3pt\hbox{$\scriptscriptstyle\frown$}}
                \over j} }_2}} \right]$ = inverse WPD of the modified coefficients
        \State $W_1 = I_1 - {{{\mathord{\buildrel{\lower3pt\hbox{$\scriptscriptstyle\frown$}}
                \over j} }_1}}$, $W_2 = I_2 - {{{\mathord{\buildrel{\lower3pt\hbox{$\scriptscriptstyle\frown$}}
                \over j} }_2}}$
    \EndProcedure
    \Procedure{Inverse ICA}{}
        \State $[II_1 \;\; II_2]^T = \mathbf{W}^{-1} [W_1 \;\; W_2]^T$
        \State $II = II_1 + II_2$
    \EndProcedure
    \Procedure{Correlation-based Decoding}{}
        \State Input $II$ and output $\left[ { \cdots \mathord{\buildrel{\lower3pt\hbox{$\scriptscriptstyle\frown$}}
                \over b}_l \cdots } \right]$
    \EndProcedure
  \end{algorithmic}
\end{algorithm}

Here, one may ask that (1) why we do not use a signal identification block to
identify which output of the ICA method is chaotic signal and which output is
jamming one, and (2) why we do not directly combine the de-jammed signals,
i.e. the ones are inputs of the inverse ICA block, and feed the outcome to
the conventional correlation-based receiver? The reason of not using a signal
identification block to select the separated chaotic signal is that the ICA method
is applied on signals that are decomposed from the single received one, and thus,
we expect that the ICA may not satisfactorily separate chaotic and jamming
signals. Therefore, if we use a signal identification block to identify chaotic signal
and remove the other one, we may loss sufficient amount of the transmitted signal.
In addition, the reason of not directly combining the de-jammed signals is that the
ICA method inherently produces the scale indeterminacy on its outputs. The scale
indeterminacy is maintained on the outputs of the WPD method and the de-jammed
signals. Consequently, directly combining the de-jammed signals may cause
destructive effects on the decoding process.

In summary, by integrating the VMD method, the proposed receiver can overcome
the limitation of a correlation-based receiver with ICA on the required number of input
signals. In addition, the proposed receiver can provide advantages of both the ICA
and the WPD methods on jamming estimation and mitigation. Moreover, in the newly
designed receiver, we do not need to identify which output of the ICA method is
chaotic or jamming signal, and thus, we do not need to send pilots together with the
chaotic signal. It means that with the new receiver, the system resources can be used
more efficiently.
\section{Simulation Results}
\begin{figure*}[!t]
  \centering
  \includegraphics[width = 15cm]{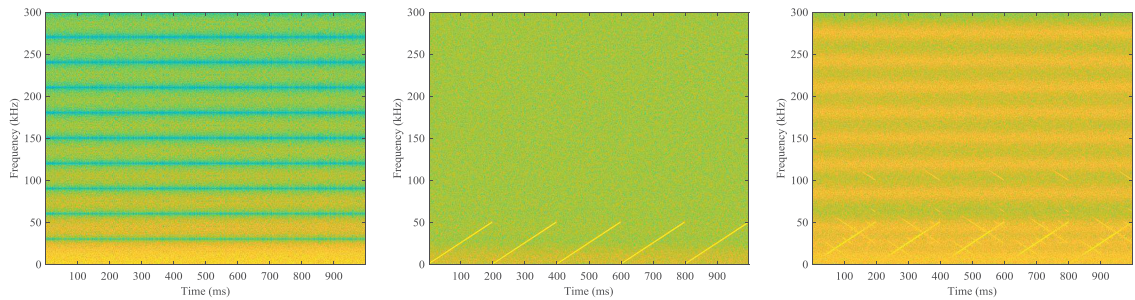}
  \caption{An example of spectrograms of the transmitted signal, received signal, and the
  signal after the inverse ICA procedure and the summation operation.}
  \vspace*{2pt}
  \hrulefill
\end{figure*}
In this section, representative simulation results are presented to show the advantages
of the proposed receiver. The simulation setting follows the system model presented
in the Section II with $\beta = 200$ and $P = 20$. In addition, the logistic generator
is used for generating chaotic samples, i.e. $x_{k+1} = 1 - 2 x_k^2$. Moreover, the
channels are modeled as block Nakagami-m fading.

In Fig. 4 shown at the top of the next page, we present an example of spectrograms of
the transmitted chaotic signal, received signal, and the signal after the inverse ICA and
the summation procedures. In the simulation, we set Eb/N0 $= 15$ dB, jamming-to-signal
power ratio (JSR) $= 5$ dB, sweep time = total transmission time/5, during which the
jammer sweeps from $0$ Hz to $50$ kHz. It can be noticed that in the received signal,
due to the jamming, fading, and noise, we can not observe the spectrum structure of the
chaotic signal. On the other hand, the jamming component can be easily noticed. In
addition, it is shown in the rightmost figure that the proposed receiver significantly
mitigates the jamming signal and preserves the spectrum structure of the chaotic one,
which suggests that the newly designed receiver can significantly enhances the AJ
performance of the NR-DCSK system.

In Fig. 5, we simulate the BER performance of the NR-DCSK system versus JSR. We
compare the system performance provided by the proposed receiver with that given by
the correlation-based and the correlation-based with WPD ones for the two cases of
Eb/N0 $= 15$ and $20$ db. It is first clearly shown that the proposed receiver provides
a similar performance to that given by the conventional counterparts in the low region of
SJR, i.e. $-20$ and $-15$ dB. However, as the SJR increases, the proposed receiver
significantly outperforms the conventional ones. More specifically, at Eb/N0 $= 20$ dB
and BER $=0.03$, the new receiver provides around $8$ dB gain compared to the
correlation-based with WPD counterpart.
\begin{figure}[!t]
    \centering
    \includegraphics[width = 7.5cm]{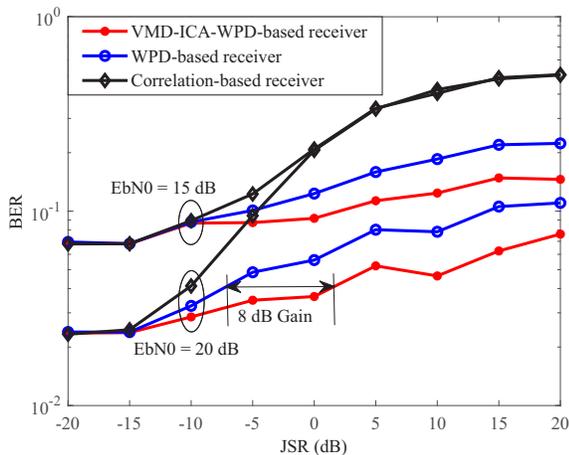}
    \caption{A comparison of the BER performances provided by the proposed,
     correlation-based, and the correlation-based with WPD receivers.}
\end{figure}
\section{Conclusions}
In this paper, we proposed a novel correlation-based receiver with VMD-ICA-WPD for the
NR-DCSK system with a single antenna destination. In the proposed receiver, the VMD method
is first used to generate multiple signals from a single received one, which alleviates the limitation
of the ICA method on the required number of input signals. Secondly, the ICA method is applied
on the outputs of the VMD method, followed by the WPD method. The WPD method is applied
on all outputs of the ICA method to estimate and mitigate jamming signals. Finally, the inverse ICA
procedure and the summation operation are carried out and the outcome is fed into the conventional
correlation-based receiver for decoding transmitted information. It is noteworthy that in the proposed
receiver, we do not need to identify which output of the ICA method is chaotic signal and which one
is jamming, from which pilots are not required and the system resources can be used more effectively.
Through extensive simulations, we showed that the newly designed receiver significantly outperforms
the conventional counterparts such as correlation-based and correlation-based with WPD receivers.
%
%
%\section*{Acknowledgment}
%The authors gratefully acknowledge the support from Electronic Warfare Research Center at Gwangju
%Institute of Science and Technology (GIST), originally funded by Defense Acquisition Program
%Administration (DAPA) and Agency for Defense Development (ADD).
%
\ifCLASSOPTIONcaptionsoff
  \newpage
\fi

\begin{thebibliography}{1}
%
\bibitem{Kaddoum-16} %2
   G. Kaddoum,
   ``Wireless Chaos-Based Communications Systems: A Comprehensive Survey,''
   {\it IEEE Access},
   vol.~4,
   pp.~\mbox{2621-2648}, May 2016.
%
\bibitem{Long-11} %4
   M. Long,
   ``Simple and Accurate Analysis of BER Performance for DCSK Chaotic Communication,''
   {\it IEEE Commun. Lett.},
   vol.~15, no.~11,
   pp.~\mbox{1175-1177}, Nov. 2011.
%
\bibitem{Xia-04} %5
   Y. Xia, C. K. Tse, and F. C. M. Lau,
   ``Performance of Differential Chaos-Shift-Keying Digital Communication Systems Over a Multipath Fading Channel with Delay Spread,''
   {\it IEEE Trans. Circuits Syst. II, Express Briefs},
   vol.~51, no.~12,
   pp.~\mbox{680-684}, Dec. 2004.
%
\bibitem{Lau-02}
   F. C. M. Lau, M. Ye, C. K. Tse, and S. F. Hau,
   ``Anti-Jamming Performance of Chaotic Digital Communication Systems,''
   {\it IEEE Trans. Circuits Syst. I, Fundam. Theory Appl.},
   vol.~49, no.~10,
   pp.~\mbox{1486-1494}, Oct. 2002.
%
\bibitem{Yang-12} %8
   H. Yang and G. P. Jiang,
   ``High-Efficiency Differential Chaos Shift Keying Scheme for Chaos-based Noncoherent Communication,''
   {\it IEEE Trans. Circuits Syst. II, Express Briefs},
   vol.~59, no.~5,
   pp.~\mbox{312-316}, May 2012.
%
\bibitem{Kaddoum-15} %9
   G. Kaddoum, E. Soujeri, C. Arcila, and K. Eshteiwi,
   ``I-DCSK: An Improved Noncoherent Communication System Architecture,''
   {\it IEEE Trans. Circuits Syst. II, Express Briefs},
   vol.~62, no.~9,
   pp.~\mbox{901-905}, Sep. 2015.
%
\bibitem{Kaddoum-16-SRDCSK} %10
   G. Kaddoum, E. Soujeri, and Y. Nijsure,
   ``Design of a Short Reference Noncoherent Chaos-based Communication Systems,''
   {\it IEEE Trans. Commun.},
   vol.~64, no.~2,
   pp.~\mbox{680-689}, Feb. 2016.
%
\bibitem{Yang-13} %7
   H. Yang and G. P. Jiang,
   ``Reference-Modulated DCSK: A Novel Chaotic Communication Scheme,''
   {\it IEEE Trans. Circuits Syst. II, Express Briefs},
   vol.~60, no.~4,
   pp.~\mbox{232-236}, April 2013.
%
\bibitem{Kaddoum-16-NRDCSK} %11
   G. Kaddoum and E. Soujeri,
   ``NR-DCSK: A Noise Reduction Differential Chaos Shif Keying System,''
   {\it IEEE Trans. Circuits Syst. II, Express Briefs},
   vol.~63, no.~7,
   pp.~\mbox{648-652}, July 2016.
%
\bibitem{Nguyen-18}
   B. V. Nguyen, H. Jung, and K. Kim,
   ``On the Anti-Jamming Performance of the NR-DCSK Systems,''
   {\it Security and Communication Networks},
   vol. 2018,
   pp.~\mbox{1-8}, Feb. 2018.
%
\bibitem{Ghatak-11}
   R. R. Ghatak, P. Sarkar, R. K. Mishra, and D. R. Poddar,
   ``A Compact UWB Bandpass Filter With Embedded SIR as Band Notch Structure,''
   {\it IEEE Microw. Compon. Lett.},
   vol. 21, no. 5,
   pp.~\mbox{261-263}, May 2011.
%
\bibitem{Regalia-10}
   P. A. Regalia,
   ``A Complex Adaptive Notch Filter,''
   {\it IEEE Signal Process. Lett.},
   vol. 17, no. 11,
   pp.~\mbox{937-940}, Nov. 2010.
%
\bibitem{Bhunia-16}
   S. Bhunia, V. Behzadan, P. A. Regis, and S. Sengupta,
   ``Adaptive Beam Nulling in Multihop Adhoc Networks Against a Jammer in Motion,''
   {\it Computer Networks},
   vol. 109, part 1,
   pp.~\mbox{50-66}, Nov. 2016.
%
\bibitem{Raju-06}
   K. Raju, T. Ristaniemi, J. Karhunen, and E. Oja,
   ``Jammer Suppression in DS-CDMA Arrays Using Independent Component Analysis,''
   {\it IEEE Trans. Wireless Commun.},
   vol. 5, no. 1,
   pp.~\mbox{77-82}, Jan. 2006.
%
\bibitem{Ranjith-16}
   J. Ranjith and Muniraj,
   ``Jammer Suppression in Spread Spectrum Communication Using Novel Independent Component Analysis Approach,''
   {\it Int. J. Electron. Commun.},
   vol. 70, no. 8,
   pp.~\mbox{998-1005}, Aug. 2016.
%
\bibitem{Mosavi-14}
   M. R. Mosavi, M. Pashaian, M. J. Rezaei, K. Mohammadi,
   ``Jamming Mitigation in Global Positioning System Receivers Using Wavelet Packet Coefficients Thresholding,''
   {\it IET Signal Process.},
   vol. 9, no. 5,
   pp.~\mbox{457-464}, Mar. 2015.
%
\bibitem{Mosavi-16}
   M. R. Mosavi, M. J. Rezaei, M. Pashaian, and M. S. Moghaddai,
   ``A Fast and Accurate Anti-Jamming System Based on Wavelet Packet Transform for GPS Receivers,''
   {\it GPS Solut.},
   vol. 21,
   pp.~\mbox{415-426}, Apr. 2016.
%
\bibitem{Nguyen-17-VMD}
   M. T. Nguyen, B. V. Nguyen, and K. Kim,
   ``Shockable Rhythum Diagnosis for Automated External Defibrillators Using a Modified Variational Mode Decomposition Technique,''
   {\it IEEE Trans. Ind. Informat.},
   vol. 13, no. 6,
   pp.~\mbox{3037-3046}, Dec. 2017.
%
\bibitem{Shen-17}
   L. Shen, Y. Yao, H. Wang, and H. Wang,
   ``ICA Based Semi-Blind Decoding Method for a Multicell Multiuser Massive MIMO Uplink System in Rician/Rayleigh Fading Channels,''
   {\it IEEE Trans. Wireless Commun.},
   vol. 16, no. 11,
   pp.~\mbox{7501-7511}, Nov. 2017.
%
%
\end{thebibliography}
\end{document}